\documentclass[aps,prb,twocolumn,groupedaddress,floatfix,showpacs]{revtex4}

\usepackage{graphics}
\usepackage{subfigure}
\usepackage{epsfig}
\usepackage{dcolumn}
\usepackage{bm}
\usepackage{overpic} 

\begin{document}

\title{The formation of Er-oxide nanoclusters in SiO$_2$ thin films with excess Si}

\author{Annett Thogersen}
\author{Jeyanthinath Mayandi}
\author{Terje Finstad}

\affiliation{Centre for Materials Science and
  Nanotechnology, University of Oslo, P.O.Box 1126 Blindern, N-0318 Oslo, Norway}

\author{Arne Olsen}
\affiliation{Department of Physics, University of Oslo, P.O.Box 1048 Blindern, N-0316 Oslo, Norway}

\author{Spyros Diplas}
\affiliation{SINTEF Materials and Chemistry, P.O.Box 124 Bildern, 0314 Oslo, Norway}

\author{Masanori Mitome}
\author{Yoshio Bando}
\affiliation{National Institute of Material Science, Tsukuba, Japan}

\date{\today}

\begin{abstract}

\noindent The nucleation, distribution and composition of erbium embedded in a SiO$_2$-Si layer were studied with High Resolution Transmission Electron Microscopy (HRTEM), Electron Energy Loss Spectroscopy (EELS), Energy Filtered TEM (EFTEM), Scanning Transmission Electron Microscopy (STEM) and X-ray Photoelectron Spectroscopy (XPS). When the SiO$_2$ layer contains small amounts of Si and Er, nanoclusters of Er-oxide are formed throughout the whole layer. Exposure of the oxide to an electron beam with 1.56*10$^6$ electrons/nm$^2$/sec. causes nanocluster growth. Initially this growth matches the Ostwald ripening model, but eventually it stagnates at a constant nanocluster radius of 2.39 nm.

\end{abstract}


\maketitle


\section{Introduction}

For a long time, silicon was considered unsuitable for optoelectronic applications because of its indirect band gap and the absence of the linear electro-optic effect. Nanocrystals of silicon doped with rare earth ions have been investigated with the prospect of potential use in optical applications, light-emitting diodes and lasers \cite{Meldrum:intro, Ennen:intro}. It was shown that doping of rare earth ions into the Si structure is beneficial for Si optoelectronic properties \cite{Zhang:intro}, especially when using erbium (Er). Er has an unfilled 4f shell surrounded by an external closed shell \cite{Zhang:intro} and intra 4f-transitions ($4l_{13/2}$ to $4l_{15/2}$) will therefore show luminescence at 1.54 $\mu$m \cite{Ennen:intro, fuji:intro}. These transitions can be excited both optically \cite{Franzo:intro} and electrically \cite{Michel:intro}.

Room temperature light emission from Er-doped Si nanocrystals has been studied extensively \cite{shin:intro, kik:intro, ce:intro, aj:intro, zhingunov:intro, xu:intro}. Most of the studies are on $\sim$300 nm thick silicon oxide films, while little attention has been given for oxide films thinner than 50 nm. On the other hand, many studies have been done on the surrounding atomic environment of Er in Si. Maurizio et al.\cite{Maurizio:intro} found that the Er atom was surrounded by only O atoms, and no direct Er-Si bonds were observed. Terrasi et al.\cite{terrasi:intro} reported that heating the sample for three hours at 620$^\circ$C created a mixed environment of Si and O around the Er atoms. Further heat treatment at 900$^\circ$C removes the residual Er-Si coordination and produces a full oxygen coordinated first shell with an average of 5 O neighbors \cite{terrasi:intro}. When the concentration of Er is high, Er forms Er-O bonds in competition with Si. Then Er$_2$O$_3$ forms in addition to ErSi$_2$ \cite{mf:intro}. These studies have been performed on Er implanted samples with high Si concentration, but few published papers present work on low Si concentration.

In previous studies the nucleation, growth and the crystal and electronic structure of Si nanoclusters in a thin SiO$_2$ layer have been studied by various microscopy and spectroscopy techniques \cite{annett:nr1,annett:nr2}. In the present work, we studied the nucleation, composition and distribution of Er clusters in SiO$_2$ with low doses of Si, by means of High Resolution Transmission Electron Microscopy (HRTEM), Electron Energy Loss Spectroscopy (EELS) mapping, Energy Filtered TEM (EFTEM), Scanning Transmission Electron Microscopy (STEM) and X-ray Photoelectron Spectroscopy (XPS).

\section{Experimental}
\noindent

The samples were made by growing a $\sim$3 nm layer of SiO$_2$ on a p-type Si substrate by Rapid Thermal Oxidation (RTO) at $1000^\circ$C for six seconds. A 30 nm layer of Si and Er-rich oxide subsequently was sputtered from SiO$_2$:Si:Er composite targets onto the initially formed RTO-SiO$_2$ film. The area propotion for Si and Er was 17 area \% (11 at. \%) and 1.1 area \% (0.1 at. \%) respectively. Sputtering was followed by heat treatment in a N$_2$ atmosphere at $1000-1100^\circ$C for 30-60 minutes.

Cross-sectional TEM samples were prepared by ion-milling using a Gatan precision ion polishing system with 5 kV gun voltage. The samples were analysed by HRTEM and EDS in a 200 keV JEOL 2010F microscope with a Gatan imaging filter and detector, and a NORAN Vantage DI+ Electron Dispersive Spectroscopy (EDS) system. When studing the nanocluster growth during electron beam exposure (see section III A), the current density was measured, as the total current density on the fluorescent screen. EFTEM, EELS mapping and STEM were performed with a 300 keV JEOL 3100FEF microscope equipped with an Omega imaging filter. EFTEM- Spectral Imaging (EFTEM-SI) was performed using energy losses from 2 eV to 30 eV and with an energy slit of 1 eV. The EELS mapping images were obtained by placing an energy slit width of 20 eV for the acquisition of pre- and postedge images around the Si L$_{2,3}$ and Er N$_{4,5}$ edges.

Elemental distribution images for Si and Er were displayed as the difference between two pre-edge images and one post-edge. A higher intensity in the experimental images reflects a higher elemental concentration. XPS was performed in a KRATOS AXIS ULTRA$^{DLD}$ using monocromated Al K$_\alpha$ radiation (h$\nu$=1486.6 eV) on plane-view samples at zero angle of emission (vertical emission) with charge neutralization. The X-ray source was operated at 10 mA and 15 kV. The spectra were peak fitted using Casa XPS \cite{casa:xps} after subtraction of a Shirley type background.

\section{Results and discussion}

\subsection{Evolution of nanocluster size}

\begin{figure}
  \begin{center}
    \includegraphics[width=0.4\textwidth]{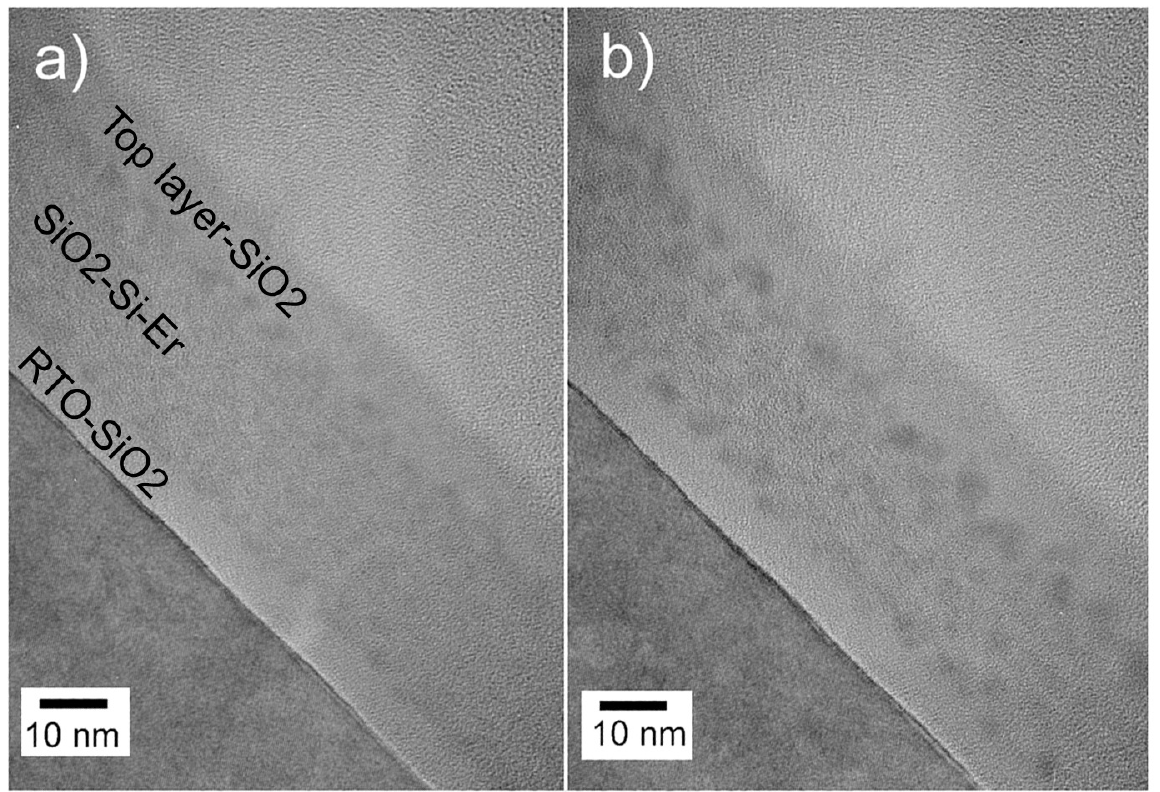}
   \caption{HRTEM images of the same area at a) an exposure time of less than 30 seconds, and b) after long exposure time to the electron beam. The images were taken with a 200 keV JEOL 2010F microscope.}
    \label{figure:1}
  \end{center}
\end{figure}

The HRTEM images taken with a 200 keV JEOL 2010F microscope, show dark areas of nanoclusters in the oxide (see Figure \ref{figure:1}). The nanoclusters were amorphous and precipitate throughout the whole oxide thickness, except for the RTO-SiO$_2$ and the SiO$_2$ top layer. At a very short electron beam exposure time (less than 20 seconds), the Si-Er-rich SiO$_2$ appears darker than pure SiO$_2$ observed in previous studies\cite{annett:nr1,annett:nr2}, and only a few small nanoclusters were observed. This is attributed to the Er atoms being evenly distributed in the oxide before exposure to an intense electron beam. Er is a heavy element and will therefore scatter more electrons. The sample was exposed to an electron beam with a current density of 39.1 pA/cm$^2$, measured on the fluorescent screen at 800 000 times magnification. The electron density on the specimen was calculated as the density on the image screen multiplied with the magnification squared, which results in an 39.1 pA/cm$^2$*800000$^2$ = 25.0 A/cm$^2$ electron density. The electron counts per nm square area for 1 second is then 1.56*10$^6$ electrons/nm$^2$/sec. After only 30 seconds of electron beam exposure at an electron density of 25.0 A/cm$^2$, the Er atoms form nanoclusters of 1.2 nm in radius. Further exposure induces nanocluster growth.

\begin{figure}
  \begin{center}
    \includegraphics[width=0.4\textwidth]{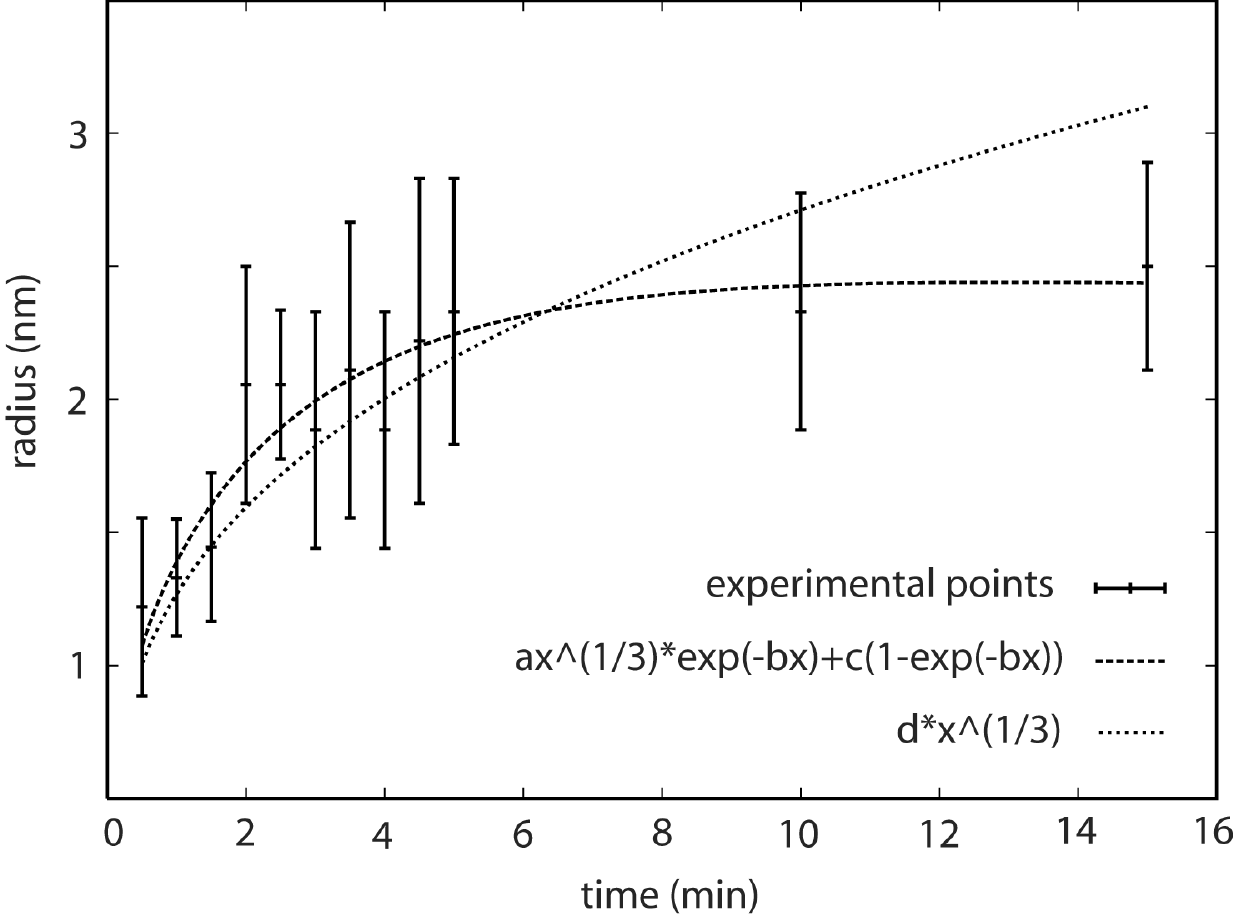}
   \caption{Figure 2: A plot of the nanocluster radius (nm) versus exposure time (min) with two fitted functions.} 
    \label{figure:2}
  \end{center}
\end{figure}

A plot of the nanocluster radius (nm) versus beam exposure time (min) is presented in Figure \ref{figure:2}. The visible nanoclusters start with a radius of 1.2 nm, with a standard deviation of $\pm$ 0.5 nm, and grow to a radius of 2.5 $\pm$ 0.5 nm over a time period of 15 minutes. After reaching 2.5 nm in radius, the nanocluster size does not change significantly.

Figure \ref{figure:2} shows measurements of particle size with time, together with two fitted curves corresponding to different models. The parameters in these models were fitted to the experimental data by the nonlinear weighted least-squares method (Marquardt-Levenberg algorithm) using the computer program Gnuplot. The weights (error-bars) are chosen to be the reciprocals of the squares of the standard deviations, so that the more precise measurements are more significant for determining these parameters.

One of the fitted functions follows the Ostwald ripening model \cite{verhoven:r, ostwald:1},

\begin{equation}
\label{eq:1}
\overline{R}^ 3=\overline{R}^3_0+[\frac{8}{9}\frac{\overline{V}^ 2 \gamma CD}{kT}]t,
\end{equation}

where R is the radius at time t, R$_0$ is the initial radius at t=0 minutes, $\overline{V}$ [cm$^3$] is the volume per atom in the nanocluster, $\gamma$ [N/m] is the surface tension, C [cm$^{-3}$] is the equilibrium concentration and T [K] the temperature. The diffusion term D [cm$^2$/s] is of the form

\begin{equation}
\label{eq:2}
D=D_0 e^{-Ea/kT},
\end{equation}

with D$_0$ [cm$^2$/s] the diffusion coefficient, $E_a$ [eV] the activation energy and k [1.38*10$^{-23}J/K$] the Boltzmann constant.

As there were only a few nanoclusters observed in the beginning of the experiment, it is reasonable to assume R$_0$ = 0 nm. Furthermore, although the sample temperature might increase slightly at the start of the experiment, it can be assumed to be more or less constant during the measurements. Thus the Ostwald ripening model can be simplified to R(t)=d*t$^{1/3}$ where d is now a new parameter.

Fitting this function as above, gives a d value of 1.27 $\pm$ 0.06 nm/min$^{1/3}$. Figure \ref{figure:2} shows that the fitted function agrees well with the experimental data at the start of the process, but does not agree with the measurements acquired at longer time. This behaviour is attributed to the decrease in Er concentration in the matrix with increasing time. Eventually there will be no Er left in the matrix, and the nanoclusters will not grow any further. To take this into account, we propose a model that initially agrees with the Ostwald ripening model, but eventually tends to a certain constant nanocrystal radius,

\begin{equation}
\label{eq:3}
R(t)=at^{1/3}e^{-bt}+c(1-e^{-bt})
\end{equation}

This is essentially a convex combination of the Ostwald ripening model and the constant function c. Fitting this function to the data as above, yields parameters a=1.21 $\pm$ 0.38 nm/min$^{1/3}$, b=0.17 $\pm$ 0.24/min and c=2.39 $\pm$0.28 nm. The nanocrystal radius therefore tends asymptotically towards c=2.39 $\pm$0.28 nm. We call this the critical nanocluster size. Below this size, the largest nanoclusters grow by Ostwald ripening at the expense of the smaller ones \cite{l:size, wagner:size, verhoven:r}. In this way, the system reduces its energy by reducing the relatively high surface potential of the smaller nanoclusters.

\subsection{Composition of nanoclusters}

EDS measurements were performed on the nanoclusters to study their composition, see Figure ref{figure:3}. Since the spectra show contributions from Er, Si and O, the nanoclusters will combine Er with either Si or O, or both.

\begin{figure}
  \begin{center}
    \includegraphics[width=0.4\textwidth]{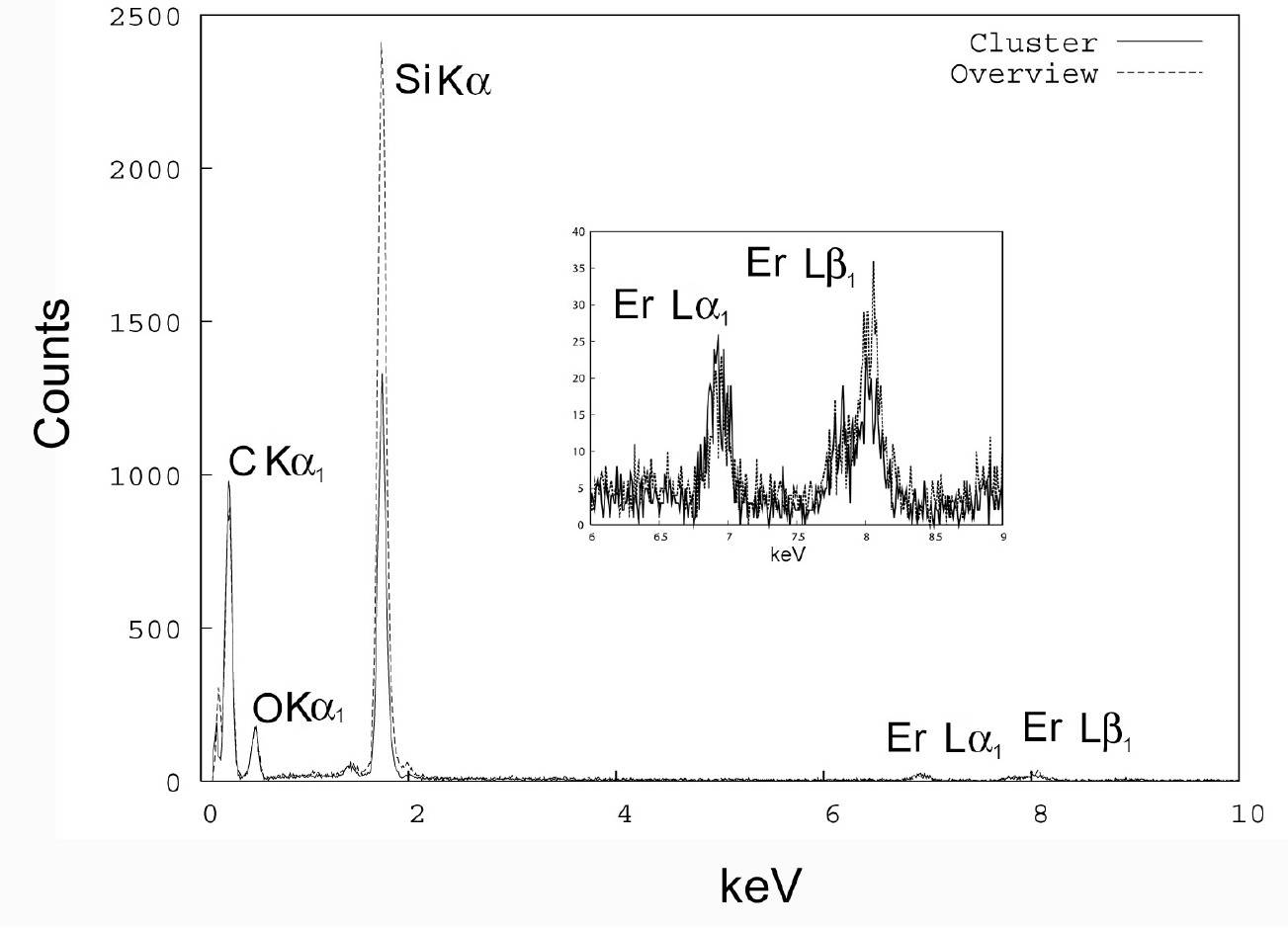}
   \caption{EDS spectrum of the nanoclusters and the oxide taken with a 200 keV JEOL 2010F microscope with a NORAN Vantage DI+ EDS system}
    \label{figure:3}
  \end{center}
\end{figure}

\begin{figure*}
  \begin{center}
    \includegraphics[width=0.7\textwidth]{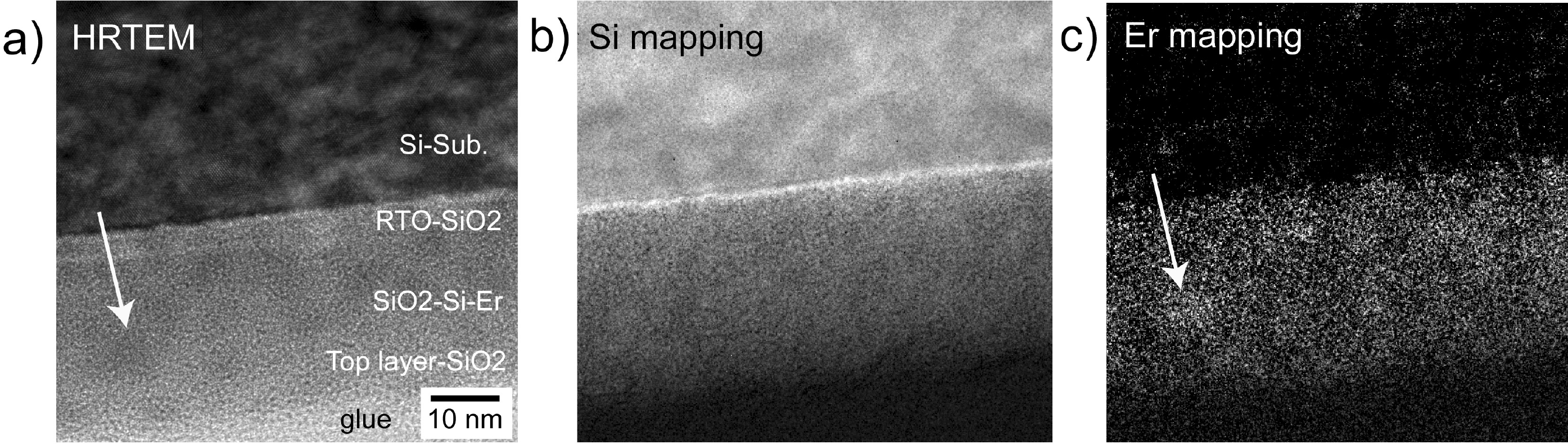}
   \caption{EDS spectrum of the nanoclusters and the oxide taken with a 200 keV JEOL 2010F microscope with a NORAN Vantage DI+ EDS system}
    \label{figure:4}
  \end{center}
\end{figure*}

\begin{figure}
  \begin{center}
    \includegraphics[width=0.4\textwidth]{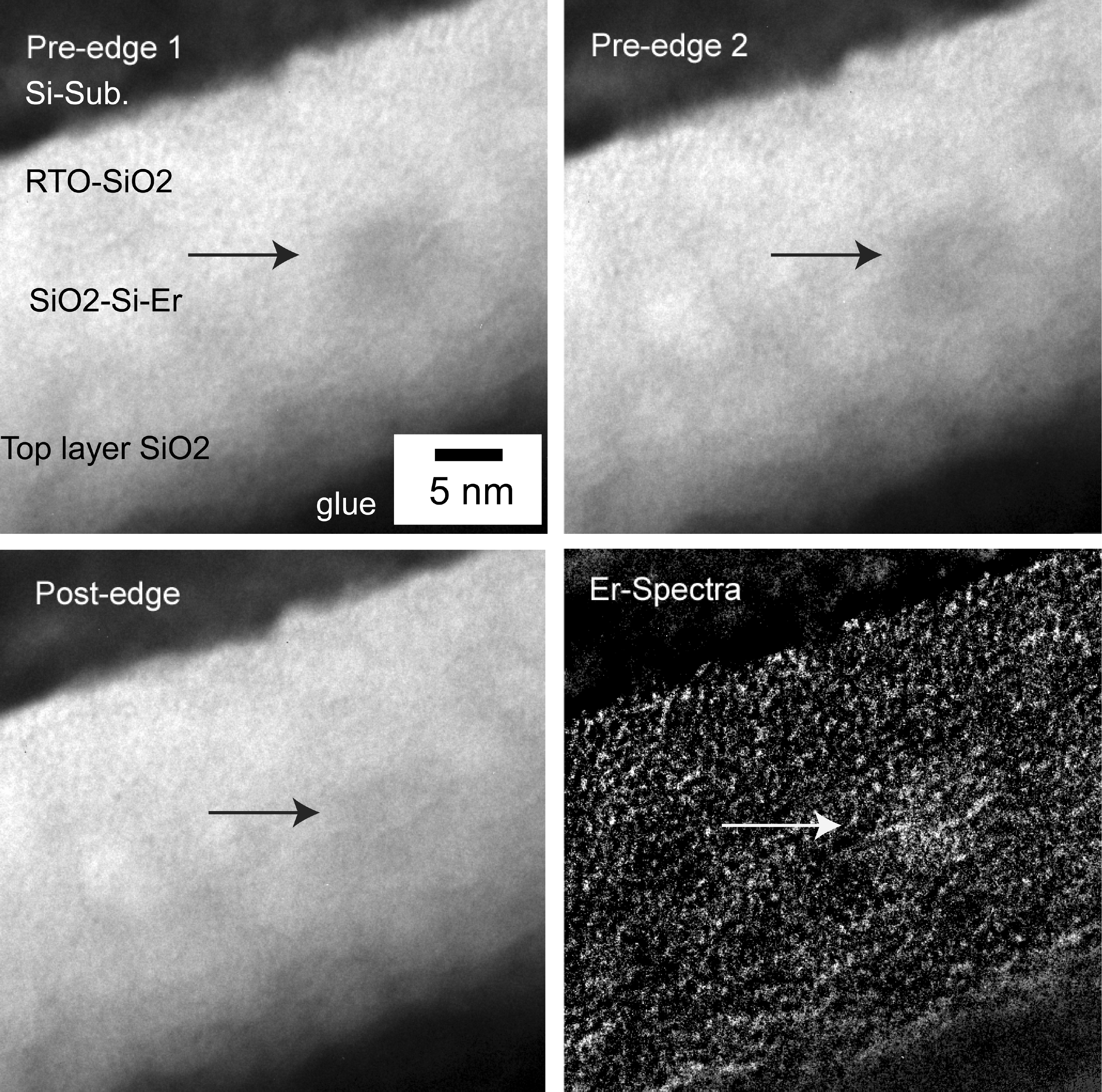}
   \caption{Two pre-edge images, one post-edge image and a mapped area of the Er N$_{4,5}$ peak. The images were acquired using a 300 keV JEOL 3100FEF microscope.}
    \label{figure:5}
  \end{center}
\end{figure}

To determine qualitatively the composition of the nanoclusters, EELS mapping, EFTEM imaging of the main plasmon peak of Si and STEM were performed on the samples using a 300 keV JEOL 3100FEF microscope. Figure \ref{figure:4}a shows an HRTEM image obtained using the contrast objective aperture as well as elemental mapping of Si (\ref{figure:4}b) and Er (\ref{figure:4}c) using the Si L$_{2,3}$ and Er N$_{4,5}$ edges. In a previous study of SiO$_2$ films containing 17 area \% Si (11 at. \% Si), very small amorphous nanoclusters of Si were detected \cite{annett:nr1}. Such nanoclusters are, however, very difficult to detect by conventional TEM imaging\cite{annett:nr2}. Figure \ref{figure:4}b shows the Si mapped image. The substrate appeares bright as expected, but no Si nanoclusters are visible in the oxide. Four Er maps of a similar area showing two pre-edge images, one post-edge image and the extracted image (post-edge - pre-edge) with higher resolution, are presented in Figure \ref{figure:5}. Er cluster is marked by an arrow. These results indicate that the dark nanoclusters seen in Figure \ref{figure:1} contain Er, in agreement with the EDS analysis.

\begin{figure}
  \begin{center}
    \includegraphics[width=0.4\textwidth]{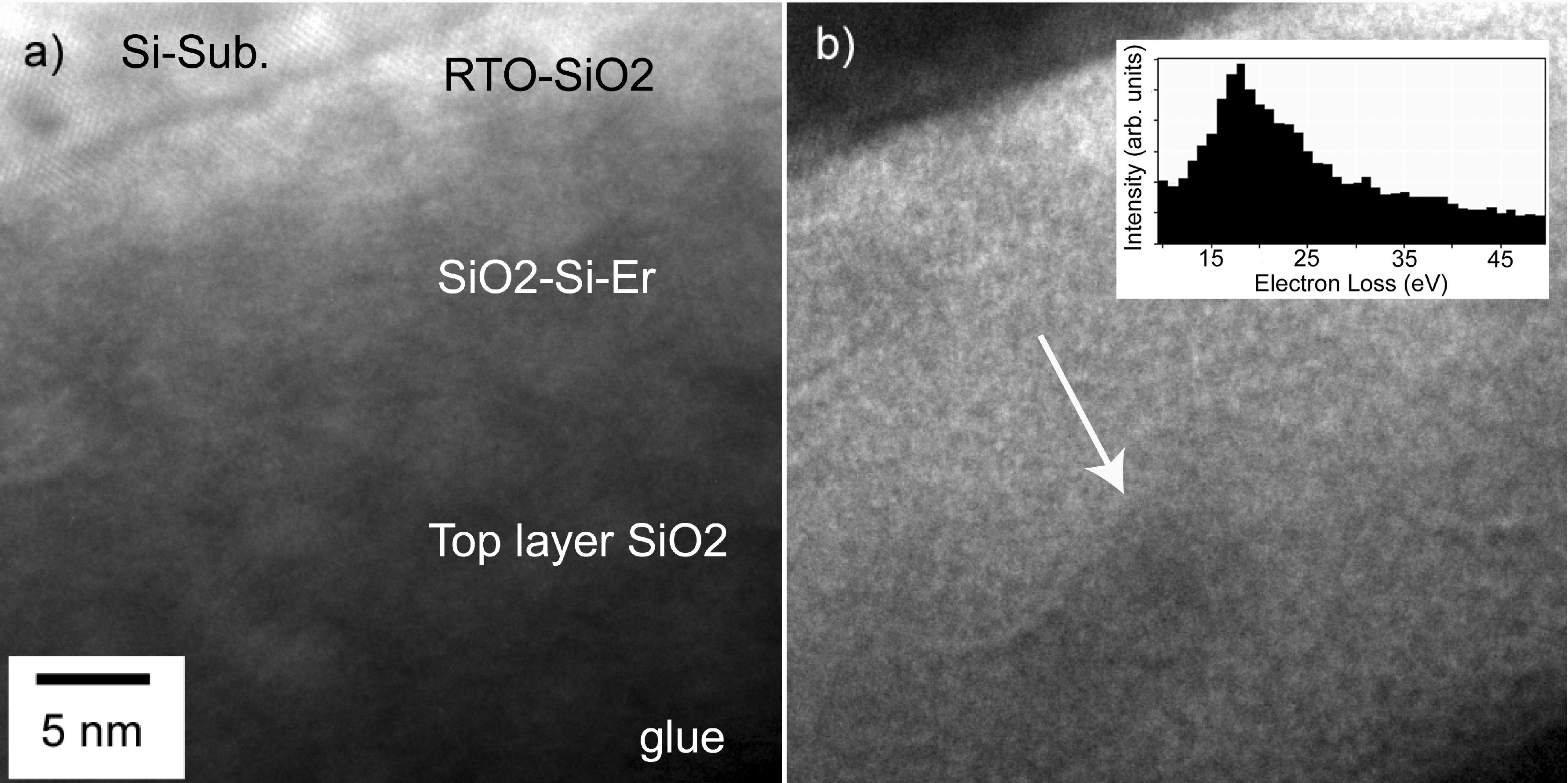}
   \caption{EFTEM image showing a) the plasmon peak of Si (16.8 eV) and b) the plasmon peak of SiO$_2$ (23 eV) taken with a 300 keV JEOL 3100FEF microscope. The inset shows an EELS spectrum of a dark nanocluster with a plasmon energy of 17 eV.}
    \label{figure:6}
  \end{center}
\end{figure}

EFTEM of the Si plasmon peak was performed to determine if the nanoclusters contain Si. Figure \ref{figure:6}a shows an image created from the plasmon peak of pure Si (16.8 eV, with 2 eV energy slit). The oxide appears dark and no Si rich nanoclusters are visible. Figure \ref{figure:6}b presents an image created from the plasmon peak of SiO$_2$ (23 eV, with 2 eV energy slit). As expected, the matrix is bright, and dark nanoclusters are visible. Therefore the dark nanocluster areas do not seem to contain any Si. The EELS spectrum inserted in Figure \ref{figure:6}b was taken from the middle of a dark nanocluster shown by the arrow. As the EELS spectrum has low energy resolution, it is not possible to extract any detailed features. The nanocluster has a plasmon peak at 17 eV. Er$_2$O$_3$ has a plasmon peak at 14 eV, and Er at 11.6 eV. Note that the nanoclusters are surrounded by SiO$_2$, and this can influence the low loss spectra, since SiO$_2$ has a plasmon energy of 23 eV and is the major contributor to plasmon oscillations. Since the plasmon energy in Figure \ref{figure:6}b at 17 eV is closer to the value for Er$_2$O$_3$ (14 eV) this indicates that the nanoclusters contain Er-oxide.

\begin{figure}
  \begin{center}
    \includegraphics[width=0.3\textwidth]{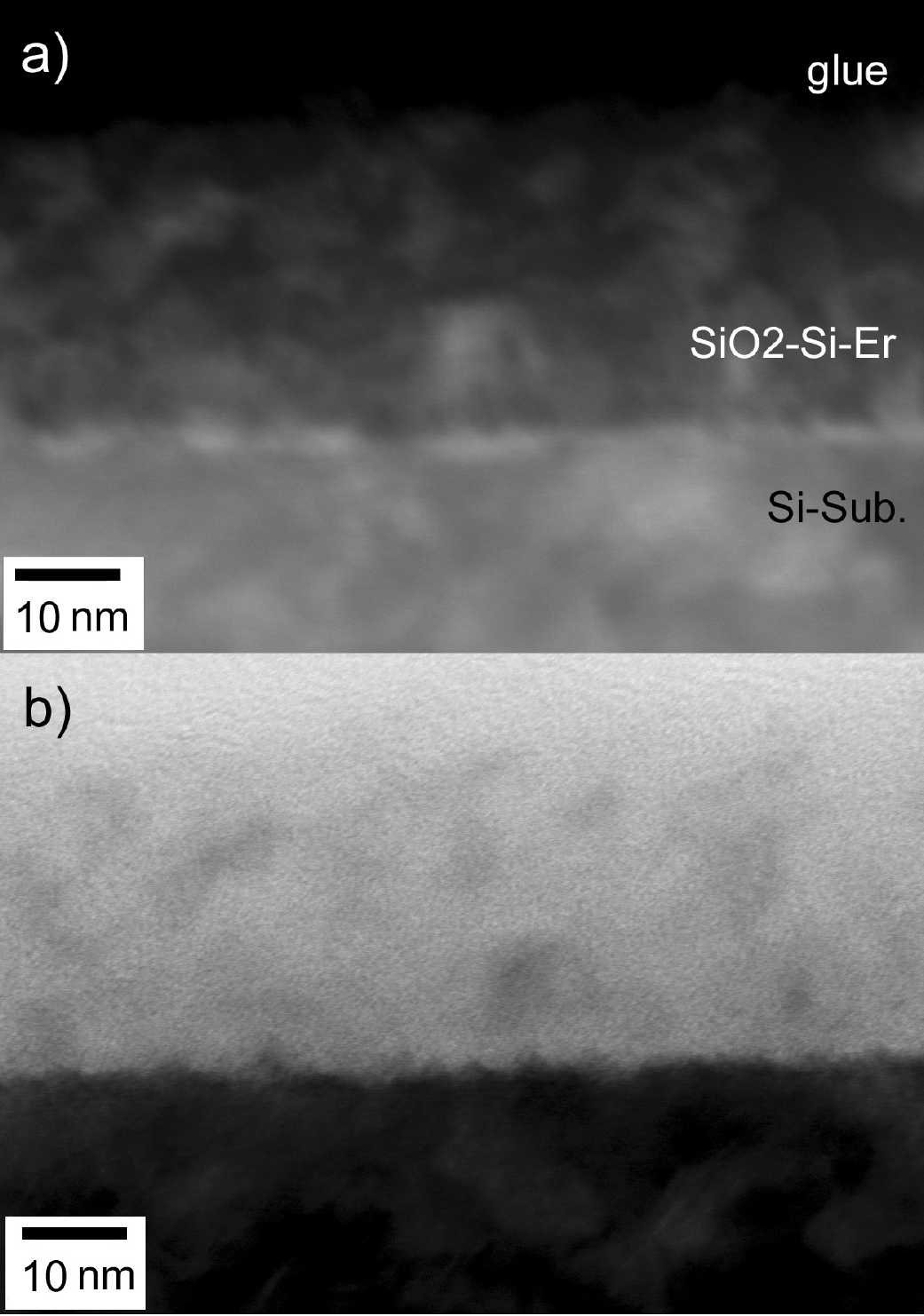}
   \caption{STEM a) HAADF image and b) BF image using a 300 keV JEOL 3100FEF microscope.}
    \label{figure:7}
  \end{center}
\end{figure}

STEM images with both Bright Field (BF) and High Angle Annular Dark Field (HAADF) detector are presented in Figure \ref{figure:7}. The STEM HAADF image in Figure \ref{figure:7}a shows bright nanoclusters of 3-5 nm in the dark oxide. In this figure there is also a thin bright area near the SiO$_2$-RTO/Si-substrate interface, suggesting the presence of a thin Er rich layer at the interface. Figure \ref{figure:7}b presents a STEM BF image of the same area as shown in Figure \ref{figure:7}a. The nanoclusters that appear bright in Figure \ref{figure:7}a are dark in Figure \ref{figure:7}b. The SiO$_2$-RTO/Si-substrate interface seems to be more irregular than that seen in Figure \ref{figure:1} and Figure \ref{figure:4}a. This could be due to additional Er precipitation at the interface during prolonged exposure to the electron beam. STEM imaging heats up the sample more than regular TEM, this could lead to diffusion of Er atoms to the SiO$_2$-RTO/Si-substrate interface. 

\subsection{Chemical state of nanoclusters}

XPS spectra of the different elements present in the sample were acquired during depth profiling with Ar$^+$ sputtering. Figure 8 shows the high resolution Er-4d spectrum at different depths in the oxide. The Er-4d spectrum from the Er rich area (spectrum seven in Figure 8) in the sample is shown in Figure 9a, after a Shirley background subtraction. This spectrum was taken from a depth well before reaching the Si substrate and therefore the signal from pure Si was absent. The Er-4d peak overlaps with the Si-2s plasmon peak of SiO$_2$. To extract the Er-4d peak from the overlapping spectrum, The Si-2p from the SiO$_2$ (Figure 9d) was used. For spectral calibration purposes the Si-2p from the Si substrate was also used (Figure 9c). Figure 9b shows the spectrum resulting from the subtraction of the SiO$_2$ plasmon peak from the overlapping Er-4d and Si-2s plasmon. The spectral positions were calibrated using the O-1s peak for SiO$_2$ at 533 eV, the Si$^{4+}$-2p peak at 103.6 eV and the Si$^0$-2p peak from the substrate at 99.5 eV \cite{xray:ref}. The subtracted spectrum in Figure 9b shows three peaks located at 170.3 eV, 175.5 eV and 184.9 eV. The literature values of the Er$_{4d}^0$ and Er$_2$O$_3$-4d binding energy are 168.0 and 168.7 eV respectively \cite{gok:xps, xray:ref}. Since EDS, EFTEM and EELS mapping showed no Si in the nanoclusters, they consist most likely of Er-O oxide. In this respect the peak at 170.3 eV arises from Er-oxide, and the peaks at 175.5 eV and 184.9 eV are energy loss peaks. 

The difference in binding energy between the reference value for bulk Er$_2$O$_3$ (168.7 eV) and the measured binding energy of the Er-4d peak (170.3 eV) may be due to quantum size effects, deviation from the Er$_2$O$_3$ stoichiometry or in energy referencing issues or a combination of all of these factors. The shift in binding energy can be expressed as

\begin{equation}
\label{eq:shift}
\Delta E_B = K \Delta q + \Delta V + \Delta \varphi - \Delta R
\end{equation}

In Equation \ref{eq:shift}, K is the Coulomb interaction between the valence and core electrons and $\Delta q$ expresses the change in the valence electrons. K$\Delta q$ describes therefore the difference in interaction between core and valence electrons. $\Delta V$ is the contribution of the changes in  Madelung potential. $\Delta \varphi$ is the changes of the sample work function which may be important in the case of insulators. These first three terms in Equation \ref{eq:shift} refer to initial state effects. The fourth term is the contribution of the relaxation energy R, which is the kinetic energy gained (negative sign in binding energy scale) when the electrons in the solid respond (screen) to the photohole produced by the photoemission process, and is a final state effect. If one could neglect relaxation effects in the above formula, an increased ionicity would lead to a higher binding energy difference ($\Delta E_B$), while an increased covalence would lead to a smaller difference.

The relaxation energy term in Equation \ref{eq:shift} reflects the core hole electron screening efficiency. Lets assume that in the case of a nanocluster besides to the contribution of the Er nanocluster itself, extra atomic screening is likely to arise from the surrounding matrix. Er$_2$O$_3$ has a higher dielectric constant\cite{diel:er2o3} than the surrounding matrix (SiO$_2$= 3.9 and Er$_2$O$_3$= 13), i.e. Er$_2$O$_3$ is more insulating than SiO$_2$. A dielectric material is a non-conducting material that can withstand a high electric field. If a material with a high dielectric constant is placed in an electric field, the magnitude of the field will be measurably reduced within the dielectric. Because of the comparatively large distance between the atoms in the dielectric, none of the atoms interact with one another. A material with high dielectric constant has low screening efficiency. If an Er-4d core hole in the Er$_2$O$_3$ nanoclusters is screened by the SiO$_2$ matrix (which has a lower dielectric constant), the screening in the nanoclusters would be superior to the screening in bulk Er$_2$O$_3$. This would be reflected as a reduction in binding energy compared to bulk Er$_2$O$_3$, however, this is not the case. The universal screening length \cite{screen:length} is expressed as

\begin{equation}
\label{eq:screenlength}
a_u = \frac{0.8854}{Z_1^{0.23}+Z_2^{0.23}} a_0
\end{equation}

where $Z$ is the atomic number for atoms 1 and 2 and $a_0$ is the Bohr radius ($52.9177*10^{-3}$nm). Calculations using equation \ref{eq:screenlength} for the Si-Si, Si-O, O-O, Er-Er, Er-O and Er-Si interactions, resulted in $a_u^{O-O}$=0.0145 nm, $a_u^{Si-O}$=0.0136 nm, $a_u^{Si-Si}$= 0.0128 nm, $a_u^{Si-Er}$=0.0105 nm, $a_u^{Er-O}$= 0.011 nm and $a_u^{Er-Er}$= 0.0089 nm. The above values are aligned with the dielectric nature of the two oxides, SiO$_2$ and Er$_2$O$_3$. The average screening length in SiO$_2$ is larger than that in Er$_2$O$_3$. However, the screening length of SiO$_2$ is very short compared to the nanocluster size, therefore, the core hole screening in the ``bulk'' of the Er-oxide nanocluster will not be largely affected by the surrounding matrix. The binding energy of the Er oxide in the nanoclusters found in this work is higher than what is reported for bulk. This increase is therefore likely to be due to initial state effects, rather than final state effects.

\begin{figure}
  \begin{center}
    \includegraphics[width=0.4\textwidth]{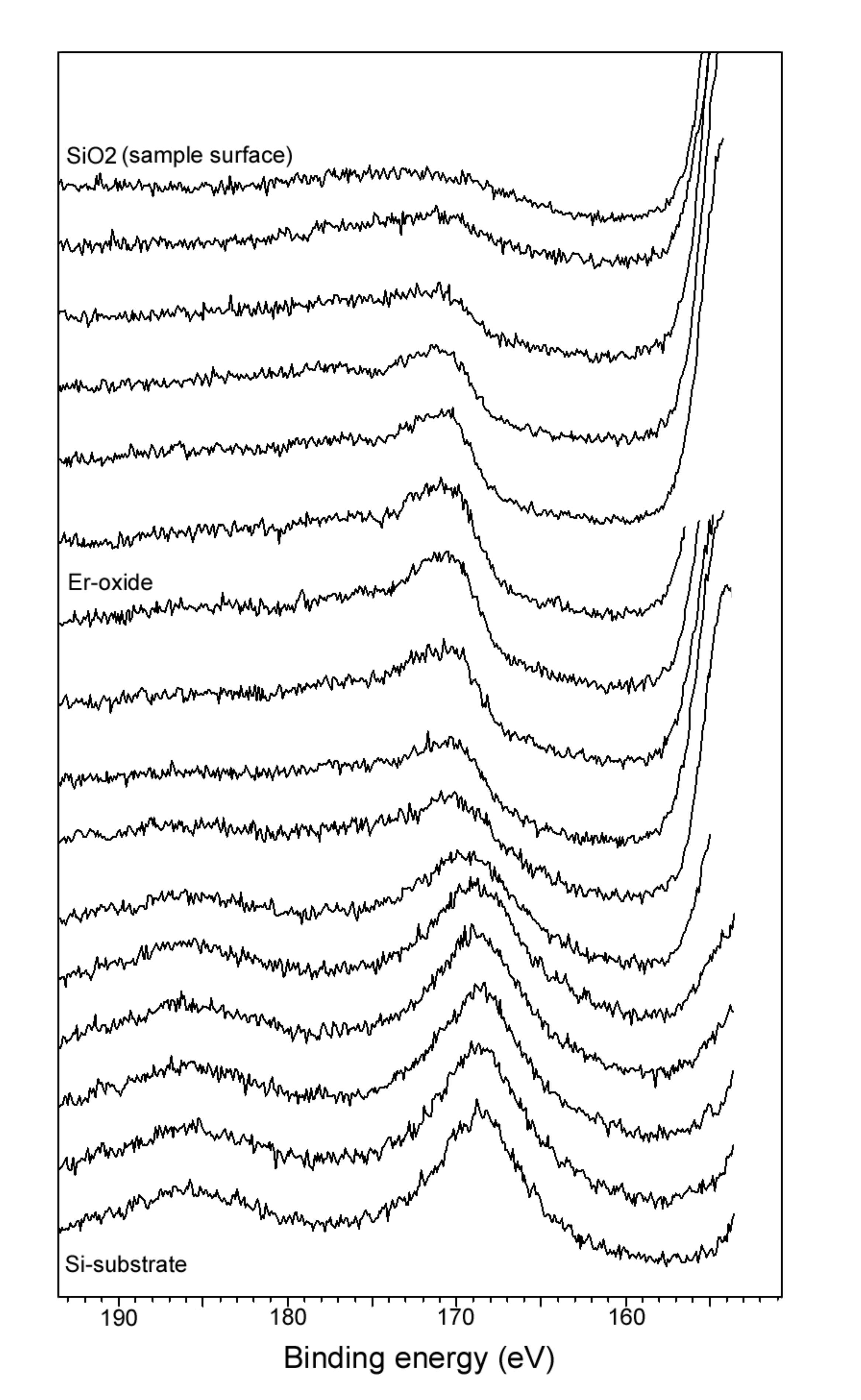}
   \caption{XPS spectra of the Er-4d peak, acquired during Ar$^+$ sputtering. The top spectrum is from SiO$_2$ and the bottom from the Si-substrate. The spectra in the middle are from Er-oxide in SiO$_2$.}
    \label{figure:8}
  \end{center}
\end{figure}

Er has a low solubility in SiO$_2$. This is due to the mismatch in size and valence between the Er ions and Si ions in SiO$_2$. In Er$_2$O$_3$, Er$^{3+}$ ions are bonded to six O atoms, with bond lengths around 0.22-0.23 nm \cite{Er:xps}. Assuming purely ionic bonding, an Er ion would donate half an electron to each of its O neighbours. A tetrahedrally coordinated Si$^{4+}$ ion , which is known to be the common Si state in amorphous SiO$_2$, would donate one electron to each O neighbour. Pure Er in SiO$_2$, in the form of ErO$_6$, fits poorly into a pure SiO$_2$ network, even when allowing for local reconstructions\cite{Er:xps}. Energy is therefore gained by forming clusters in which several Er ions can share nearest neighbours \cite{Er:xps}.

The valence spd levels of Er are higher in energy than the valence sp levels of Si, with respect to the oxygen valence band level. Charge transfer from Er to O will therefore be favoured over charge transfer from Si to O. The amount of charge transfer from Er to O will depend on the amount of charge that O recieves from the other neighbours (Si in this case). The replacement of Si for Er in the next-nearest-neighbour shell of an Er atom will therefore increase the charge transfer away from the Er sites in accordance with numerical results (since Si to O charge transfer is less favourable)\cite{Er:xps}. Laensgaard calculated the charge transfer (in units of e) for Er atoms in several compounds, with crystalline Er$_2$Si$_2$O$_7$ and Er$_2$O$_3$ amongst them. It was found that Er in Er$_2$Si$_2$O$_7$ shows a charge transfer of 0.290$e^-$, while Er in Er$_2$O$_3$ shows a charge transfer of 0.284-0.276$e^-$, a positive number means that electrons are transferred away from the atom. Guittet et al. \cite{Er:xps2} found the charge transfer for Si and O in SiO$_2$ to be 2.05$e^-$ and -1.02$e^-$ respectively.

In the case of Er-oxide nanoclusters, initial state effects seem to play a profound role as the cluster size decreases. In the context of the above discussions, at the interface between Er$_2$O$_3$ nanoclusters and SiO$_2$ matrix, charge transfer from Er towards O is expected to be larger as compared to the ``bulk'' of the nanoclusters. This additional charge transfer from the Er sites, $\Delta q$ may be the reason for the increased binding energy of the Er-4d peak in the nanoclusters as compared to bulk Er$_2$O$_3$. The Pauling electronegativity values for Si, O and Er are 1.9, 3.44 and 1.24 respectively. These values support the argument for an increased charge transfer from the Er sites towards the O neighbours in the presence of Si.

It is noticeable that the binding energy shifts in the case of Er$_2$O$_3$ nanoclusters occur in the opposite direction of those of amorphous Si nanoclusters in a SiO$_2$ matrix\cite{annett:nr1}. In both cases it seems that initial state effects govern the shifts.

\begin{figure}
  \begin{center}
    \includegraphics[width=0.4\textwidth]{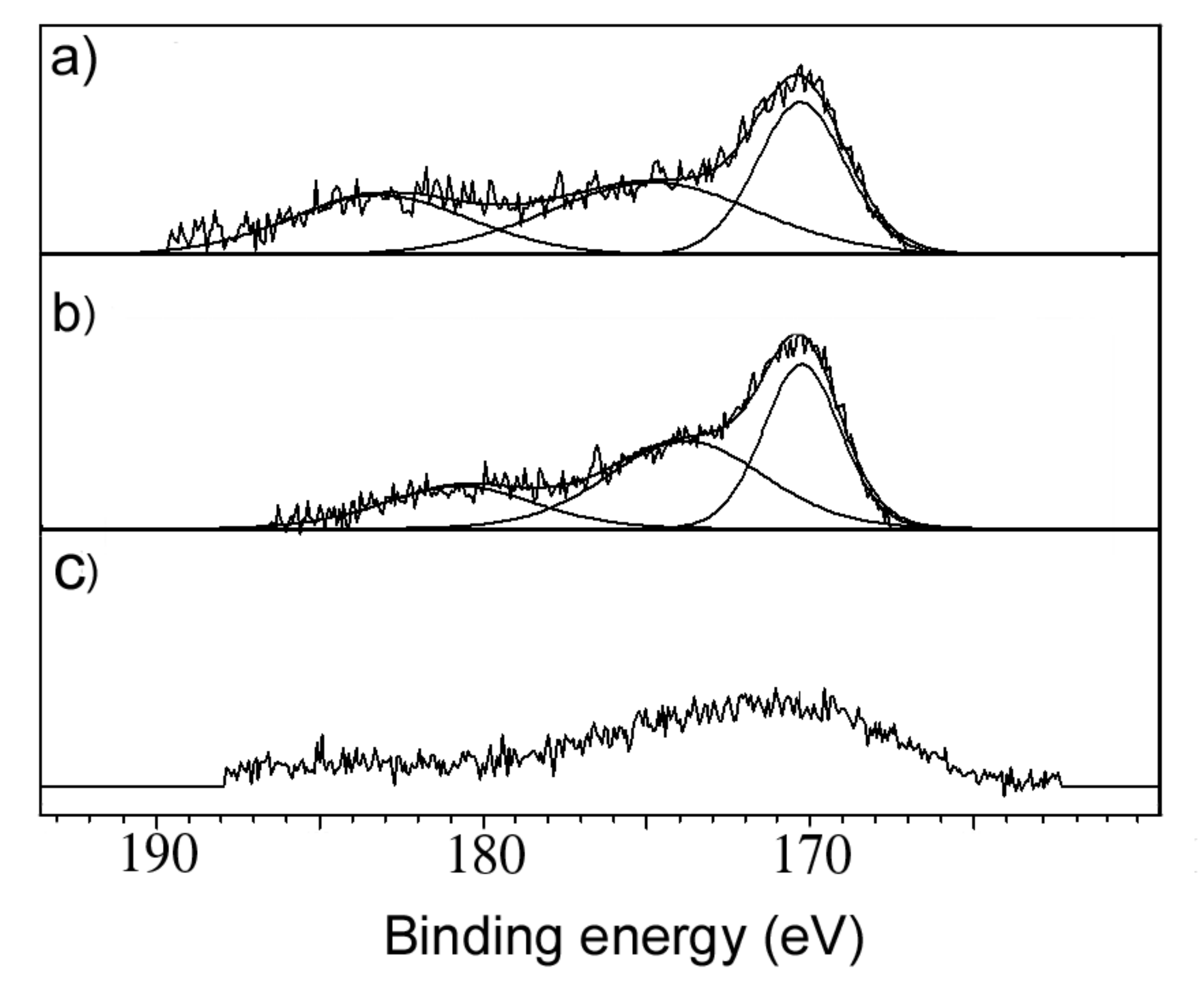}
   \caption{XPS spectrum of a) the Er-4d peak overlapping with the plasmon peak of SiO$_2, b) $the Er-4d peak (after subtracting the SiO$_2$-2s plasmon spectrum), c) the SiO$_2$-2s plasmon peaks.}
    \label{figure:9}
  \end{center}
\end{figure}

\section{Conclusion}

When the SiO$_2$-Si-Er layer contains low Si concentrations, Er-oxide (possible Er$_2$O$_3$) clusters nucleate throughout the oxide. Under the exposure to an electron beam with 1.56*10$^6$ electrons/nm$^2$/sec., the nanocluster radius grows initially according to the Ostwald ripening model, but eventually grows asymptotically towards a nanocluster radius of 2.39 nm. The increased Er-4d binding energy of Er-oxide in nanoclusters apart from energy referencing reasons could be attributed to initial state effects dominated by increased charge transfer from Er towards O in the presence of Si at the nanocluster-SiO$_2$ matrix interface.

\section{Acknowledgement}

\noindent Financial support by FUNMAT@UiO, the University of Oslo, Kristine Bonnevie's and SCANDEM's travelling scholarship is gratefully acknowledged. One of the autors (A.T.) would like to thank Georg Muntingh for many helpful comments and discussions.

\end{document}